\documentclass[12pt]{article}

\usepackage{scicite}
\usepackage{times}
\usepackage{aas_macros}
\usepackage{graphicx}
\usepackage{xcolor}
\usepackage{authblk}
\usepackage{amssymb}
\usepackage{xspace}
\usepackage{ulem}

\newcommand{\sv}[1]{\textcolor{black}{#1}}
\newcommand{\msun}{\ensuremath{M_\odot}\xspace}
\newcommand{\chieff}{\ensuremath{\chi_{\mathrm{eff}}}\xspace}

 \newcommand{\muG}{\ensuremath{\mu_{\mathrm{G} }= 33.49^{+4.54}_{-5.51}}\xspace}
 \newcommand{\mninetyninewith}{\ensuremath{m_{99\%, \mathrm{BH}} = 59.6^{+14.5}_{-16.1}   }\xspace}
 \newcommand{\mninetyninewithout}{\ensuremath{m_{99\%,  \mathrm{BH}} = 52.1^{+13.8}_{-7.3}   }\xspace}
 \newcommand{\monewith}{\ensuremath{m_{1\%,  \mathrm{BH}}= 2.5^{+0.3}_{-0.3}   }\xspace}
 \newcommand{\monewithout}{\ensuremath{m_{1\%,  \mathrm{BH}}= 6.0^{+1.0}_{-1.6} }\xspace}
 \newcommand{\alphaLVC}{\ensuremath{ \alpha=- 2.62^{+0.62}_{-0.73}  }\xspace}
  \newcommand{\sigmaG}{\ensuremath{ \sigma_{\mathrm{G}}=5.09^{+4.28}_{-4.34}  }\xspace}
  \newcommand{\GaussFrac}{\ensuremath{10^{+14}_{-7}  }\xspace}
  \newcommand{\chimu}{\ensuremath{\mu_\mathrm{eff}=0.06^{+0.05}_{-0.05}}\xspace}
 \newcommand{\fracnegchi}{\ensuremath{27^{+17}_{-15}}\xspace}
\newcommand{\FinalMassMayEvent}{\ensuremath{142^{+28}_{-16}}\xspace}

\newenvironment{sciabstract}{%
\begin{quote} \bf}
{\end{quote}}

\title{The first five years of gravitational-wave astrophysics}

\author
{Salvatore Vitale$^{1,2}\footnote{\normalsize{$^\ast$To whom correspondence should be addressed; E-mail: salvatore.vitale@ligo.org}}$ \\
\small{$^{1}$Laser Interferometer Gravitational-Wave Observatory Laboratory, Massachusetts Institute of Technology, Cambridge, 02139, USA}\\
\small{$^{2}$Department of Physics and Kavli Institute for Astrophysics and Space Research, Massachusetts Institute of Technology, Cambridge, 02139, USA}\\
}

\date{\today}
\begin{document}
\baselineskip24pt
\maketitle
\begin{sciabstract}
Gravitational waves are ripples in spacetime generated by the acceleration of astrophysical objects. A direct consequence of general relativity, they were first directly observed in 2015 by the twin Laser Interferometer Gravitational-Wave Observatory (LIGO) observatories. I review the first five years of gravitational wave detections. More than fifty gravitational waves events have been found, emitted by pairs of merging compact objects such as neutron stars and black holes. These signals yield insights into the formation of compact objects and their progenitor stars, enable stringent tests of general relativity and constrain the behavior of matter at densities higher than an atomic nucleus. Mergers that emit both gravitational and electromagnetic waves probe the formation of short gamma ray bursts, the nucleosynthesis of heavy elements, and measure the local expansion rate of the Universe.
\end{sciabstract}
\clearpage

Albert Einstein's general theory of relativity (GR), published in {1915}~\cite{Einstein1915}, showed that gravity is not a force, but an effect of spacetime curvature on anything with mass-energy; that is, on everything. Mass and energy deform spacetime and hence {are} the sources of gravity. This interplay between sources and geometry was described by John Archibald Wheeler as ``Spacetime tells matter how to move; matter tells spacetime how to curve''~\cite{Misner2018}.

GR predicts the existence of gravitational waves (GWs)~\cite{Einstein1918}: ripples in the fabric of spacetime generated by masses as they accelerate. In Newtonian gravity changes in the gravitational potential propagate instantaneously, but the GWs described by GR travel at the speed of light.
The existence of wave-like solutions is a direct mathematical consequence of Einstein's field equations, but it was long unclear whether GWs are physically measurable even in principle~\cite{Kennefick}.
In the 1950s it was established that a gravitational-wave train would deposit energy in a detector, and thus was potentially measurable.
This led to the development of resonant-bar GW detectors in the 1960s, which were used to report  detections in 1969 and 1970~\cite{Weber1969}, but those results could not be verified or replicated by independent researchers.

\begin{figure}
\centering
\includegraphics[width=1\textwidth]{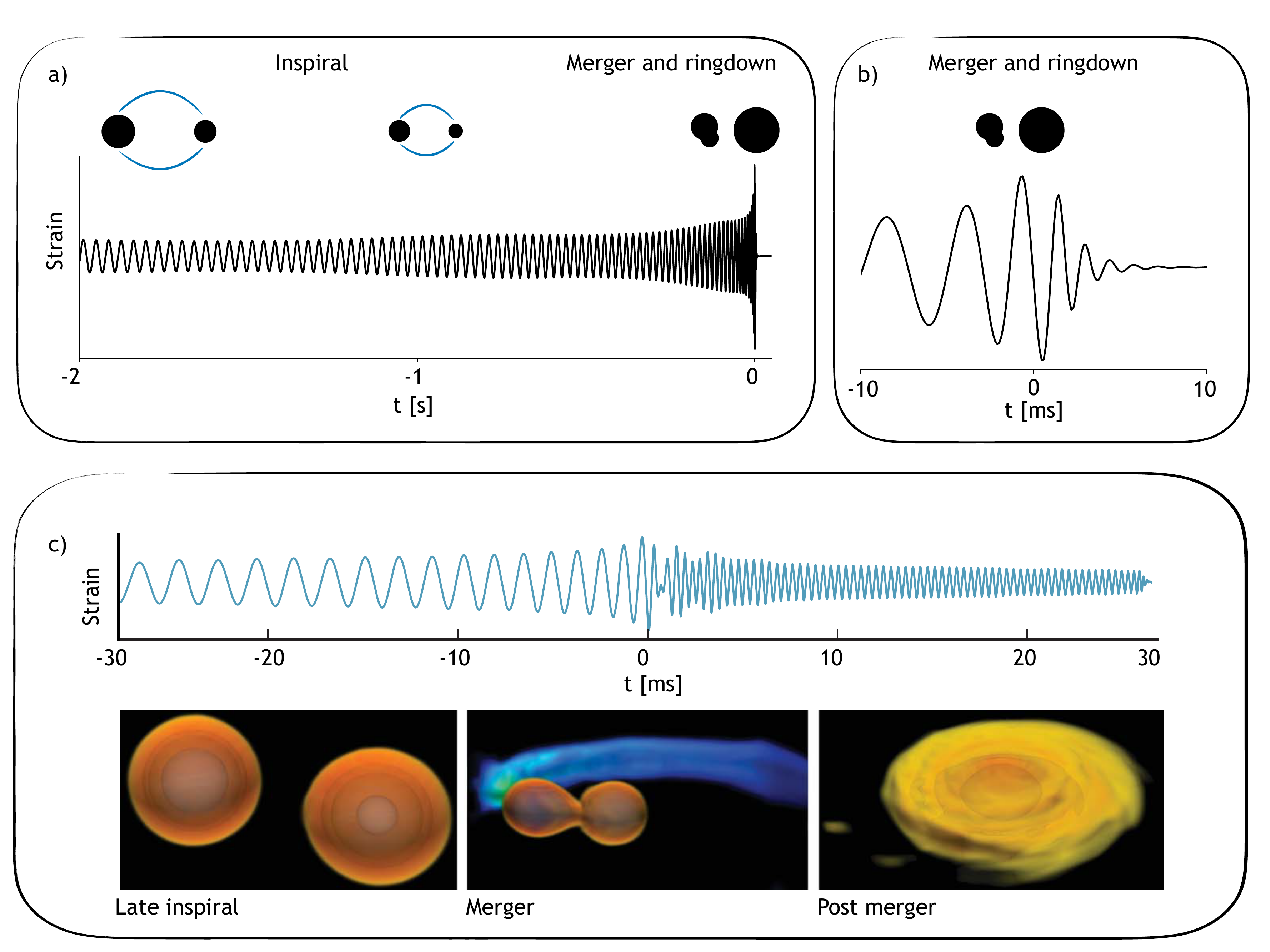}
\caption{\textbf{Gravitational waveforms from compact binaries.} a) The predicted final two seconds of the gravitational waveform produced by a binary black hole, with diagrams of the two compact objects. The amplitude modulation in the inspiral phase is due to spin-induced precession of the orbital plane b) The same signal in the 10 milliseconds around the merger time. After the merger, the product black hole rings down to an equilibrium state. c)  The predicted gravitational-wave signal in the 30 milliseconds around the merger time of a binary neutron star merger. High-frequency GW emission continues after the merger. The neutron stars are initially tidally deformed and then disrupted in the late inspiral. A disk of matter forms around the final compact object. The color reports the density of the material, from low (blue) to high (dark red) (Credits for panel c, T. Dietrich, based on numerical simulations of Ref.~\cite{Dietrich:2015iva}). The numerical value of the waveforms amplitude depends on the sources parameters, including their distance, and is not reported.
\label{Fig.Combo}}
\end{figure}

Conclusive evidence for the existence of GWs came indirectly. GR predicts that two objects in a binary system will lose energy through emission of GWs. This energy loss shrinks the orbit, enhancing the emission of GWs, producing a runaway process that ends when the two objects collide (Fig.~\ref{Fig.Combo}).
The discovery of a binary neutron star system in 1974, with one of the two objects visible in the radio band as a pulsar, allowed this prediction to be tested. Data collected over the next 8 years showed that the orbit was shrinking as predicted for GW emission~\cite{1982ApJ...253..908T}. Another prediction of GR, the existence of black holes, was verified in the same period: by the mid 1970s most astronomers were convinced that the X-ray source in Cygnus X-1 was most likely a black hole of roughly 15 solar masses (\msun). Black hole binary systems would emit even more GW energy than neutron star binaries, providing  potential sources of GWs that  would be easier to detect.

\section*{Interferometric detectors}

Detecting GWs requires monitoring the distance between freely falling objects. GWs are quadrupole in nature, implying that objects lying along perpendicular directions are affected by the waves in opposite ways: e.g. distances decrease along the {x} direction and increase along the {y} direction.
The principle of detecting this effect is straightforward, but the technological hurdles appeared insurmountable in the 1960s: adopting plausible values for the strength of potential GW sources gave a relative displacement induced by GWs -- and hence the required precision of measurement -- of $\sim10^{-21}$ in the audio frequency band.

By the early 1970s, it was realised that this could potentially be achieved using laser interferometry~\cite{Weiss1972}. After another 3 decades of research and development, 
the Laser Interferometer Gravitational-wave Observatory (LIGO) began taking data in 2002. LIGO, consists of two laser interferometer detectors in Louisiana and Washington United States. This was followed by the  Italian-French detector Virgo in 2007. 

These three detectors took data until 2010, without reporting any detections. The detectors were then shut down so they could be upgraded, with the goal of providing a factor of ten improvement in sensitivity.

\begin{figure}[t]
\centering
\includegraphics[width=0.75\textwidth]{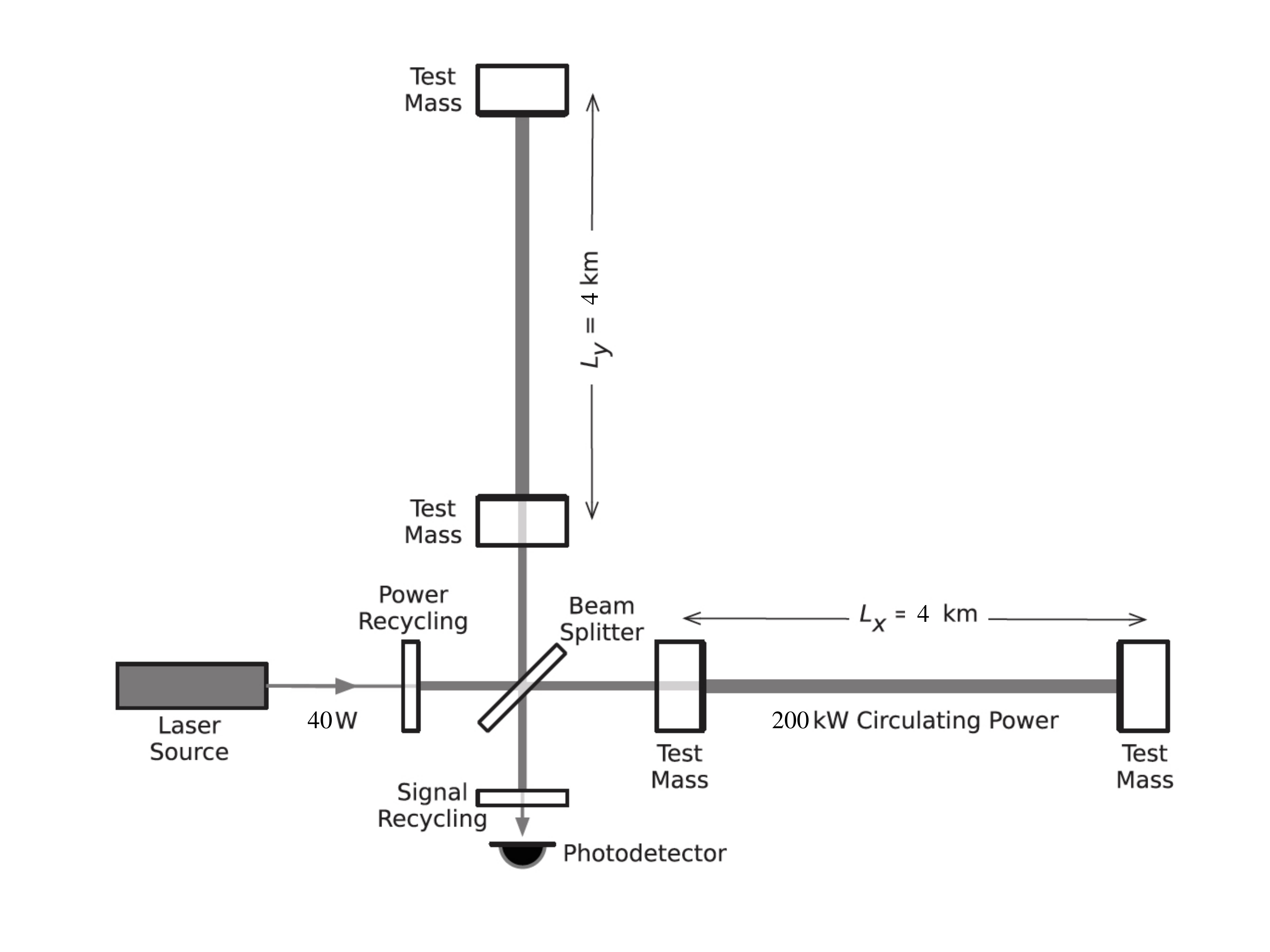}
\caption{\textbf{A schematic of the Advanced LIGO detector.} Light from the laser source goes through a beam splitter and enters the 4 kilometer long orthogonal arms. Power and signal recycling mirrors are used to increase the sensitivity of the detector. Adapted from Ref.~\cite{Abbott:2016blz}}\label{Fig.Ifo}
\end{figure}

Fig.~\ref{Fig.Ifo} shows a schematic of the Advanced LIGO detectors~\cite{2015CQGra..32g4001L}. The light from a \mbox{high-power} laser is split and sent along the two perpendicular arms of a Michelson interferometer each arm forming a Fabry-Perot cavity. Each arm contains two mirrors which act as test masses, seismically isolated from the ground by 4-stage pendulum suspensions, as well as active and passive isolation systems. The transmittivity and reflectivity of the mirrors are chosen such that photons bounce back and forth $\sim 300$ times, before being recombined on a photodetector. Additional power recycling and signal recycling mirrors are used to increase the sensitivity of the instrument. The detectors are operated such that, in the absence of a gravitational-wave signal, light from the orthogonal arms interferes (nearly, but not exactly) destructively at the photodetector. With a 4 km arm length, detecting a relative length variation of $\sim10^{-22}$ requires measuring the position of the mirrors to better than the size of a proton. This precision is achieved by averaging over the many atoms of the mirror and the numerous photons of the laser beam.
The Advanced Virgo detector~\cite{Acernese2015} works in a similar manner, with some different design choices (such as the suspension of the mirrors), and a  shorter 3 km arm length.

\section*{Direct detection of gravitational waves}

The Advanced LIGO detectors were scheduled to start their first observing run on 18 September 2015~\cite{Aasi:2013wya}. However, a few days before, on September 14 2015, a signal was detected by both LIGO detectors during an engineering run preceding the observing run. Subsequent analysis excluded the possibility that the signal was unintentionally or maliciously added into the data stream. Nothing that could have caused the signal was found in the auxiliary channels that monitor e.g. seismic activity, winds and cosmic rays. It was a real gravitational wave, detected, 99 years after Einstein derived the equations that govern them. 

The source was designed GW150914, after the date it was detected. It was identified as the merger of two black holes, each of roughly 30~\msun, colliding at half of the speed of light to form a $62$~\msun black hole, at a distance of $\sim400$ mega parsecs (Mpc). Binary black hole systems had not previously been observed~\cite{Abbott:2016blz}. Black holes as massive as 30~\msun had not been expected. Decades of observations following the discovery of Cygnus X-1 had shown that black holes in X-ray binaries had masses between roughly 5 and 15~\msun, whereas the supermassive black holes found at the center of most galaxies had masses of millions or billions of solar masses. 
If the component black holes of GW150914 were the end-product of the lives of massive stars, then their progenitors must have been born in an environment very poor in metals~\cite{2016ApJ...818L..22A} (in astronomy any element heavier than helium is referred to as a metal). Metal-rich stars lose more of their initial mass during their lifetimes, owing to higher opacities and hence higher stellar winds, so are expected to leave behind lighter black holes.

Two more binary black hole (BBH) mergers, made of lighter component objects, were detected in the first observing run, which concluded in January 2016. This showed that the prompt detection of GW150914 was not a fortunate accident~\cite{TheLIGOScientific:2016pea}. In fact, the BBH merger rate inferred from these three events, together with the expected increase in sensitivity of the advanced detectors~\cite{Aasi:2013wya}, implied that LIGO and Virgo would detect many BBHs in their following observing runs.

\begin{figure}[t]
\centering
\includegraphics[width=0.75\textwidth]{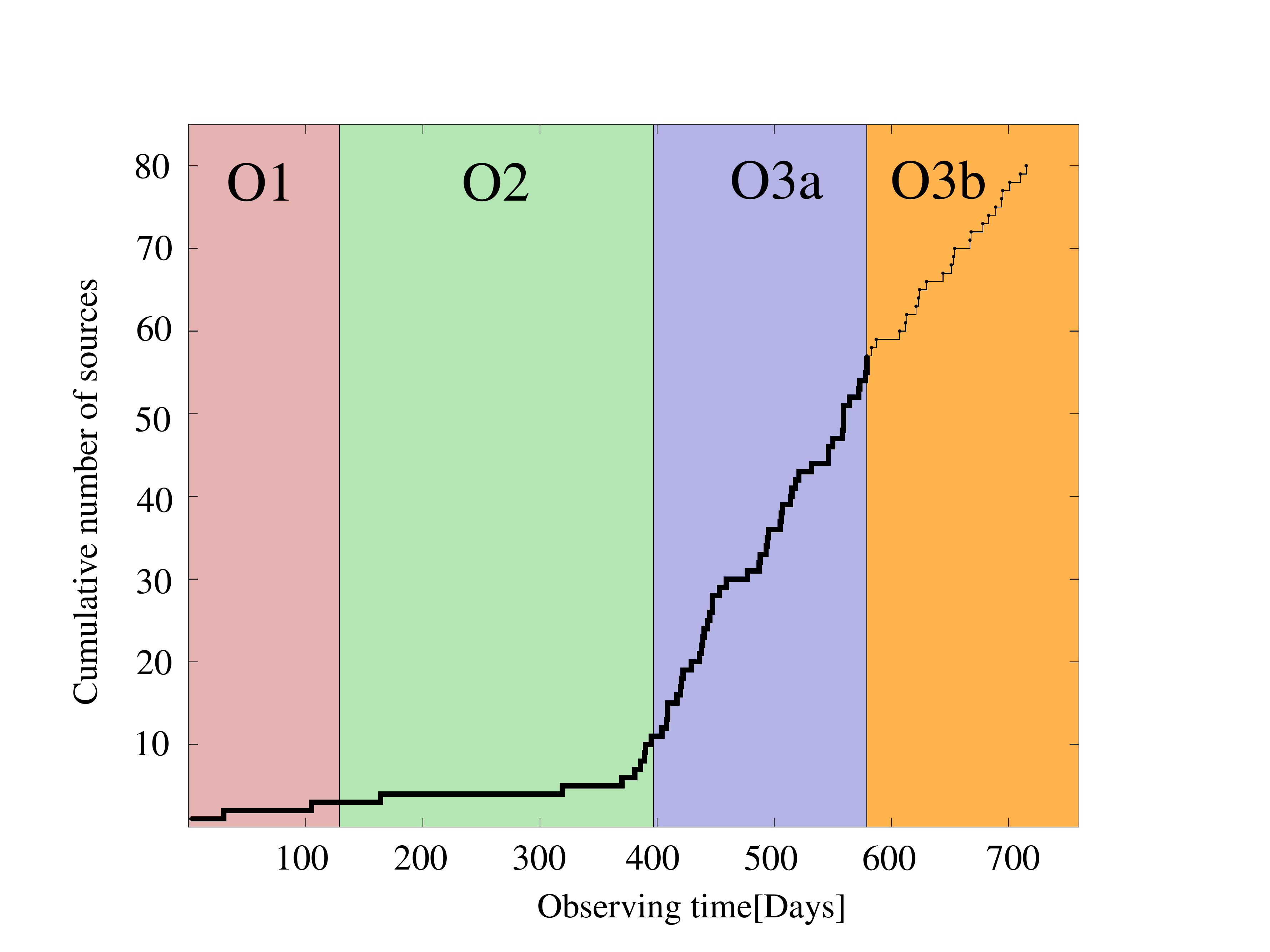}
\caption{\textbf{Gravitational-wave sources detected to date.} The cumulative number of detections (thick line) and non-retracted public alerts (thin line) in the first three observing runs (the third observing run, O3, is split in two parts, O3a and O3b). The significant change of slope between O2 and O3 is due to the increased sensitivity of the detectors\label{Fig.Detections}. (Adapted from Ref.~\cite{Reitze}).}%
\end{figure}

In the two following observing runs -- designated O2 (2016-2017) and O3 (2019-2020) -- the number of detections has continued to grow. Fig.~\ref{Fig.Detections} shows the cumulative number of confirmed and candidate GW sources that the LIGO-Virgo Collaboration (LVC) has found in the first three observing runs. The 
 change of slope between O2 and O3 was driven by improvements in the sensitivity of the instruments, including the implementation of quantum squeezing of the laser light in LIGO~\cite{Tse2019} and Virgo~\cite{Acernese:2019sbr}.
 
The LVC catalog published in \sv{April} 2021 reports 39 sources~\cite{O3aCat} found in the first half of O3 (O3a), most of which are BBHs. Additional detections, usually with low signal-to-noise ratios, have been reported by other groups, using public data from the first two observing runs~\cite{Venumadhav2020,Zackay2019,Nitz2019,Nitz2020}.

\section*{Theoretical origin of stellar mass black holes}

The detection of massive BBHs using GW data raises the issue of how such systems formed.
Theoretically, the two most likely avenues to form compact binary systems are isolated formation in low density regions such as galactic fields~\cite{1976ApJ...207..574S,Kalogera2007} and dynamical encounters in dense environments such as globular clusters~\cite{Phinney1991,Benacquista2013}.
In the first scenario, the progenitors of the two black holes are two massive stars in a stellar binary. The more massive star evolves until it increases in radius, becoming a giant that fills its Roche lobe (the region within which material can remain gravitationally bound to the star). This causes a common envelope of gas to form around both stars, which shrinks the binary's orbit through dynamical friction. After both stars run out of fuel  and explode as supernovae, the two black holes are left in a tight orbit, close enough to merge within the age of the Universe. The two black holes in binaries formed this way are expected to have similar masses, and spin vectors nearly aligned with the orbital angular momentum~\cite{1976ApJ...207..574S,Kalogera2007}. 

The alternative dynamical formation scenario instead assumes that the two black holes formed separately in a region that is densely populated by black holes. In this case, the the probability of randomly meeting a partner and forming a bound binary is non-negligeable. Globular star clusters have been proposed as the most suitable location for this to occur, as stars near the center are separated by distances similar to the size of the Solar System. As the stars in the cluster evolve, black holes form after a few million years and migrate toward the center of the cluster through dynamical friction. At sufficiently high densities, two-body and three-body interactions become frequent enough to form binaries. The black holes in a binary formed through this channel are also expected to have similar masses, but their spin vectors should be randomly oriented, owing to the random orientation of the encounter~\cite{Phinney1991,Benacquista2013}.

The mass distribution of stars heavier than the Sun is well described by a power law, with a probability density function $p(m)\sim m^\alpha$,  and a slope $\alpha \in [-2.7,-2.3]$ ~\cite{Kroupa2002}. At a given metal content, it is expected that the masses of the black holes they generate have the same functional form. However, the positions of both the lower and upper ends of this distribution are not known, and are the subject of current research.

Theoretical predictions based on nuclear physics suggest that the relation between progenitor mass and black hole mass is not monotonic. If the progenitor star mass is in the range  $\sim [100-130]$~\msun, a runaway production of electron-positron pairs at the core triggers a series of contractions and expansions which eject the outer layers of the star. These stars are expected to produce a supernova that leaves behind a black hole of $\lesssim 50$~\msun, irrespective of the progenitor mass. If the progenitor is larger than $\sim 130$~\msun but smaller than $\sim 250$~\msun, the explosion triggered by the first compression is so violent that it entirely disrupts the star, leaving no black hole behind. For stars more massive than $\sim 250$~\msun, nothing halts the gravitational collapse and a black hole is formed without any explosion: the resulting black hole will be as massive as the core of the progenitor. If these models are correct, there should exist a pair-instability supernovae (PISN) ``mass gap'' in the mass range $\sim [50-150]$~\msun, where black holes cannot form from stellar collapse~\cite{Woosley:2016hmi}.

At the other end of the mass spectrum, black holes observed in X-ray binaries have masses larger than $\sim 5$~\msun, which provides some  evidence for a mass gap between neutron stars -- which are as massive as $\sim 2$~\msun -- and stellar mass black holes~\cite{1998ApJ...499..367B,2011ApJ...741..103F}. If this gap exists, it could provide information about the physics of supernova explosions and stellar binary evolution~\cite{2010ApJ...725.1918O}.

\section*{The black hole mass distribution}

GW detections of BBH mergers~\cite{LVC02Pop,O3aCat} have been used to test these theoretical expectations.

\begin{figure}[t]
\centering
\includegraphics[width=0.85\textwidth]{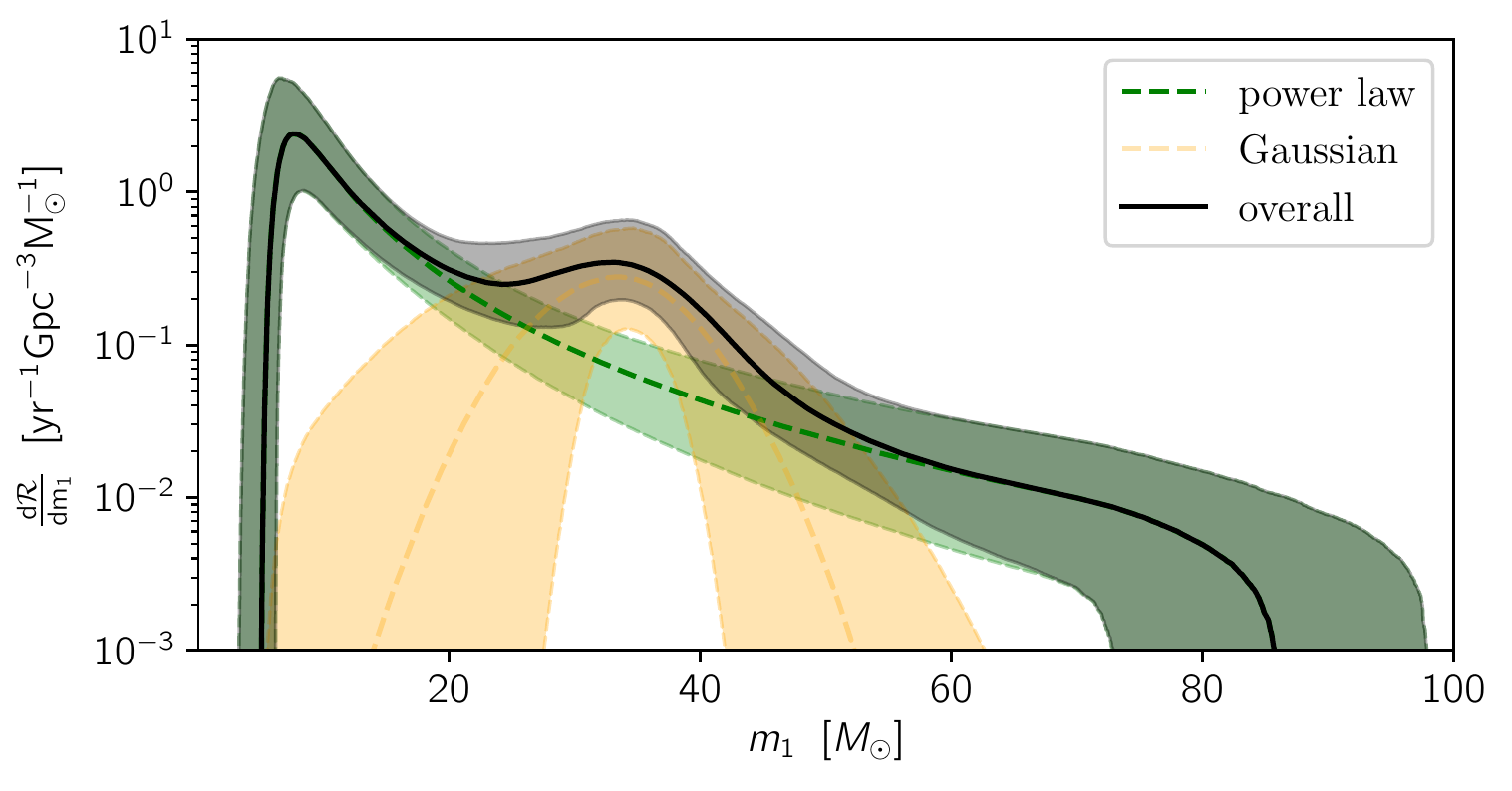}
\caption{\textbf{The black hole mass distribution.} Posterior probability distribution (gray) of the primary (i.e. most massive) black hole mass distribution, expressed as a differential merger rate $\mathcal{R}$ (with units of mergers per year per Gpc$^3$) per unit primary mass $m_1$ (with units of \msun)~\cite{O3aPop}. This is the sum of a power law (green) and a Gaussian component (orange), includes GW190521, and accounts for selection effects. The thick lines represent the medians, whereas the shaded areas are the 90\% credible intervals~\cite{O3aPop}. }\label{Fig.Mass}
\end{figure}

Many analytical models for the mass distribution of black holes in binaries have been considered~\cite{O3aPop}, with different levels of complexity. The most favoured model~\cite{O3aPop} describes the mass distribution of the primary (i.e. most massive) black hole in the detected binaries as the sum of a power-law distribution and a Gaussian distribution (Unless otherwise stated, I adopt the results from the ``power-law + peak'' analysis of Ref.~\cite{O3aPop} which excludes GW190814.). The power-law component represents the black holes that are generated by the death of massive stars,
whereas the Gaussian component describes the black holes that might pile up just below the pair-instability gap. This model has 8 free parameters, including the minimum and maximum mass generated by the power-law component and its slope, the mean and standard deviation of the Gaussian component and the fraction of BBHs that belong to it. The resulting posterior probability distribution of the primary mass distribution, and the contributions from the two components, are shown in Fig.~\ref{Fig.Mass}.

The measured slope of the  power law component -- \alphaLVC (median and 90\% credible intervals unless stated otherwise) -- is consistent with that of the parent stellar population. In most detected BBHs, the two components have similar masses, consistent with expectations for the main formation channels described above.
The Gaussian component has a mean \muG~\msun and standard deviation \sigmaG~\msun, and it contains \GaussFrac~\% of the BBHs. Models that do not include a Gaussian component are disfavored by the data, with odds ratios $\sim 4$ or worse depending on the specific alternative model.
Using the model favored by the data, the 99$^\mathrm{th}$ percentile of the mass distribution of the underlying black hole population is \mninetyninewith~\msun.

The determination of the maximum mass is largely driven by GW190521~\cite{May21Det}. With primary mass $m_1=90.9^{+29.1}_{-17.3}$~\msun and secondary mass $m_2=66.3^{+19.3}_{-20.3}$~\msun, GW190521 is the heaviest BBH found. Both components are consistent with being in the PISN mass gap. The black hole formed by the merger has a mass of \FinalMassMayEvent, consistent with being an intermediate-mass black holes (IMBHs), the theorised link between stellar mass black holes and the supermassive black holes found at the centers of most galaxies~\cite{2020ARAA..58..257G}.

Although not required by the GW data, it is possible that one or both of the component objects of GW190521 were in turn the result of a previous merger~\cite{May21Astro}, which could explain their unusually high masses. Dynamical formation in regions with a high density of black holes  could produce multiple rounds of mergers. If the black hole formed by a first-generation merger has a lower recoil velocity than the escape velocity of its environment, it is retained and can merge again. This might happen more easily in high escape-velocity environments, such dense star clusters and the disks of gas around supermassive black holes.
The repeated merger of IMBHs could lead to the formation of supermassive black holes of billions of solar masses, which are observed to have already existed when the Universe was only a few hundred million years old~\cite{2020ARAA..58..257G}. 
If GW190521 is the result of a second-generation merger, or otherwise an outlier, then it is justified to remove it while analyzing the properties of the rest of the BBH population: doing so yields \mninetyninewithout~\msun~\cite{O3aPop}, close to the expected lower boundary of the putative PISN mass gap~\cite{Woosley:2016hmi}.

The measurement of the lower edge of the black-hole mass distribution is similarly driven by one source, GW190814~\cite{LVCGW190814}. GW190814 was produced by the merger of a $23.2^{+1.1}_{-1.0}$~\msun black hole with a $2.6^{+0.1}_{-0.1}$~\msun compact object. This latter object is either the lightest black hole or the heaviest neutron star known. If GW190814 is the merger of a black hole and a neutron star, then it would be the first discovered. Unfortunately, it is hard to be sure:
owing to the large mass ratio of GW190814, even if the lighter object was a neutron star, it would not be tidally deformed before crossing the event horizon of its companion. That implies that the merger would not be accompanied by  electromagnetic (EM) emission (see discussion below).
The unconstrained tidal deformation of the lighter object, and the lack of EM counterparts, do not determine the nature of the lighter object. A $2.6^{+0.1}_{-0.1}$~\msun neutron star is possibly too heavy to be supported by the (poorly constrained) equation of state of nuclear matter~\cite{LVCGW190814}.
If both objects in GW190814 are black holes, the 1$^\mathrm{st}$ percentile of the black hole mass distribution is \monewith~\msun, narrowing or eliminating the putative mass gap between neutron stars and black holes. Alternatively, if GW190814 is treated as a neutron star-black hole merger, so excluded from the BBH analysis, \monewithout~\msun, consistent with observations of X-ray binaries~\cite{O3aPop}.

\section*{Constraints on black hole spins}

The spins of black holes in binaries also carry information about their formation pathways.  Unfortunately, measurement of the component spins is usually less precise than that of the component masses~\cite{Vitale2014,Purrer2016}. 

Astrophysical inferences can be drawn using auxiliary parameters: such as \chieff~\cite{2008PhRvD..78d4021R}, a mass-weighed projection of the total spin along the orbital angular momentum, which is usually measured much more precisely than either component spin. 
\chieff is defined in the range between $-1$ (both black holes have the largest spin allowed by general relativity and the spins are oriented along the negative direction of the orbital angular momentum) and $+1$ (same as $-1$, but  the spins are aligned along the positive direction of the orbital angular momentum). A binary can have zero \chieff either because both black holes have no spins, or because their spins are perpendicular to the orbital angular momentum.

For BBHs formed in isolated stellar binaries, the individual spins should be preferentially aligned with the orbital angular momentum, so \chieff should be positive, whereas BBHs formed dynamically have random spin orientations, meaning that \chieff can have both signs. 

While the results from LIGO-Virgo's first two observing runs were consistent with a distribution of \chieff centered at zero\cite{Note}, the additional sources from O3a suggest a different scenario. 
The measured~\cite{O3aPop} median for the \chieff distribution is \chimu, indicating that not all of the black holes have zero spin, though spins are typically small, {or along the orbital plane}. Some studies suggested that most black holes are born with negligible spins~\cite{Fuller2019}.
\fracnegchi~\% of the BBHs have negative \chieff~\cite{O3aPop}, which might indicate that some of them are formed dynamically~\cite{Rodriguez:2016avt}.

The level of misalignment between the orbital angular momentum and the component spins remains to be measured precisely, but existing results show that a configuration where all of the spins are aligned with the orbital angular momentum is excluded at 99\% credibility~\cite{O3aPop}. Misalignment between the spin and the orbital angular momentum results in orbital precession, which induces a characteristic phase and amplitude modulation in the GW signal~\cite{Apostolatos1994}, see Fig.~\ref{Fig.Combo}A. However, the evidence for misalignment in the current dataset is driven by weak unresolved features in many sources, not large precession in a few of them.

GW190412 has unusual spin parameters~\cite{2020arXiv200408342T}. The dimensionless spin magnitude -- defined in the range between 0 (no spin) and 1 (maximally spinning) -- of the most massive black hole in the binary is large, $a_1 \in [0.4,0.6]$ (depending on the specific waveform model used), and some evidence exists for orbital precession.The source's  large mass ratio, $\sim 4:1$, is unlike most other detected BBHs~\cite{LVC02Pop,2019MNRAS.484.4216R,O3aCat}, leading to competing interpretations of its origin, e.g.~\cite{Rodriguez2020,Mandel2020,2020arXiv200504243G,2020ApJ...900..177K}.

\section*{Electromagnetic counterparts}
Detecting GWs and light from the same source, a form of multi-messenger astrophysics, provides substantial additional information.
In a binary neutron star (BNS) merger, the strong tidal pull of the companion is expected to deform and disrupt each object, generating a tidal tail of debris which travels at $\sim 20\%$ of the speed of light. Shortly after the merger, an accretion disk forms around the newly created compact object (Fig.~\ref{Fig.Combo}C). BNS mergers have long been proposed as the progenitors of short gamma ray bursts (sGRBs), short ($\lesssim 2$ seconds) and energetic ($\sim10^{52}$~ergs) flashes of gamma rays from distant galaxies. sGRBs are powered by matter accretion from the disk onto the central compact object, forming a jet along the direction of the magnetic field that emits a narrow beam of gamma rays~\cite{1984SvAL...10..177B}.

As the gamma ray jet loses energy by interacting with the interstellar medium, the beaming angle increases and the electromagnetic (EM) emission moves to lower energies, and hence lower frequencies~\cite{Berger2014}. Meanwhile, the neutron-rich material surrounding the compact object (referred to as mass ejecta) is predicted to produce large amounts of EM radiation, generated by the decay of heavy nuclei, forming a {kilonova} visible in the optical band for hours to a day following  the merger~\cite{Metzger2017}. 

All of these predictions were verified with GW170817~\cite{Abbott2017GW170817,Abbott:2018wiz}, the first BNS merger detected with GWs.
The GW signal was first found by a real-time search pipeline as a BNS candidate in the LIGO Hanford data only, due to the presence of a transient instrumental noise artifact at around the time of the event in the LIGO Livingston data and issues with the transfer of Virgo data. 
The {Fermi} Gamma-ray Space Telescope had already (six minutes earlier) issued an alert reporting the discovery of an sGRB roughly 1.7 seconds after the merger time of the putative BNS event~\cite{2017ApJ...848L..13A}.
Five hours later, after having cleaned the LIGO Livingston data and having acquired Virgo data, the LVC released an updated sky map of the possible merger locations. The BNS sky map was entirely contained within the GRB sky map constraining the location of the source to an ellipse of $\sim 31 \mathrm{deg}^2$, at a distance of about 40~Mpc~\cite{MMA,Abbott2017GW170817}.

Numerous ground- and space-based telescopes started searching that patch of sky, and less than 11 hours after the merger time an optical source was identified on the outskirts of the galaxy NGC 4993, which was not present in previous observations~\cite{MMA}.
Extensive follow-up detected emission in the ultraviolet, infrared, X-rays and radio~\cite{MMA,Coulter2017,2018ApJ...856L..18M,2017ApJ...848L..20M,2017ApJ...848L..17C,2017Sci...358.1565E}, and subsequent analyses showed that the evolution of the EM emission was unlike that of any other known astrophysical transient~\cite{2017ApJ...848L..16S}. This was the first observation of light and gravitational waves from the same source.
GW signals from BNSs are morphologically richer than those from black hole binaries~\cite{Lehner:2001wq}, see Fig.~\ref{Fig.Combo}. In the late phase of the inspiral, tidal deformation deposits energy into the neutron stars, affecting the evolution of the gravitational wave~\cite{2008PhRvD..77b1502F}. The compact object formed by the collision can be a stable neutron star or an unstable neutron star that will collapse into a black hole after a few {milliseconds to seconds}; alternatively it could collapse directly into a black hole~\cite{2015PhRvD..91f4001T}. Each possible outcome results in a different set of observables in the post-merger gravitational-wave signal~\cite{2019ApJ...880L..15M} (as well as in the EM emission).
These effects depend on the unknown equation of state (EOS) of nuclear matter~\cite{1998PhRvC..58.1804A}.
This implies that compact binary mergers can be used to constrain the EOS of neutron stars, complementing EM-based methods (e.g. using the Neutron Star Interior Composition ExploreR - NICER - X-ray timing experiment~\cite{2012SPIE.8443E..13G}).

Tidal effects are larger at high frequencies (above a few hundred Hertz) where the sensitivity of LIGO and Virgo is low. Therefore, unless additional constraints are imposed, only an upper bound can be set on the tidal deformability -- a parameter that depends on the EOS and quantifies how easy or hard it is to compress a neutron star --  of the compact objects in GW170817~\cite{Abbott2017GW170817,Abbott:2018wiz}. The GW data alone cannot rule out both objects having zero tidal deformability, the value expected for black holes. Nevertheless, the EM counterparts and the estimated compact object masses are strongly suggestive of a BNS merger.
The measurement of the tidal parameters can be improved {by assuming} that both objects are neutron stars, governed by the same equation of state. This yields $190^{+390}_{-120}$ for the dimensionless tidal deformability of a $1.4$~\msun neutron star~\cite{Abbott2018EOS}, which disfavors large neutron star radii~\cite{2016ARA&A..54..401O}. Empirical relations between the tidal deformability, the masses, and the radius~\cite{Yagi2017}, have been used to convert the constraint on the tidal deformability into a measurement of the radii of the two compact objects in GW170817, $11.8^{+2.7}_{-3.3}$~km and $10.8^{+2.9}_{-3.0}$~km~\cite{Abbott2018EOS}.
The measurements can be improved by requiring that the EOS supports NS masses as large as $1.97$~\msun, a conservative estimate for the mass of the heaviest NS known at the time of the analysis, obtaining $11.9^{+1.4}_{-1.4}$~km for both objects~\cite{Abbott2018EOS} (the two equal radii is derived from the analysis not assumed ). These results can be expressed as the pressure inside the  neutron star, finding  $p(2 \rho_{\mathrm{nuc}})= 3.5^{+2.7}_{-1.7}\times 10^{34}$ dyn cm$^{-2}$ for the pressure $p$ at twice the nuclear saturation density $ \rho_{\mathrm{nuc}}$~\cite{Abbott2018EOS}. 

The multimessenger observations of GW170817 have allowed the constraints on tidal deformability to be improved by exploiting EM information. Modeling the time-varying spectral energy distribution of the EM emission provides an estimate of the amount of ejecta produced during the merger. That, in turn, depends on physical properties of the merger, such as the neutron star masses and the tidal deformability. By requiring that enough ejecta is produced to match the observed EM emission, a lower limit for the tidal deformability has been obtained~\cite{2018ApJ...852L..29R}.
Further improved constraints have been obtained by combining GW and NICER data~\cite{Raaijmakers:2019dks,2019ApJ...887L..24M}, and accounting for a newly discovered pulsar with a mass likely above $2$~\msun~\cite{2020NatAs...4...72C}.

Numerous additional results have been derived from the EM follow-up campaign of GW170817 (e.g. \cite{Kasliwal2017}). The observations have shown that at least some sGRBs are produced by merging BNSs and provided insights into the geometry of the relativistic jet~\cite{2017ApJ...848L..13A}. They have shown that BNS mergers generate kilonova emission, whose description might require two or more components, with different energetics and neutron-fractions~\cite{2017Sci...358.1565E}. In turn, this implies that a large fraction of the chemical elements heavier than iron are produced during BNS mergers, rather than SNe explosions~\cite{2017Sci...358.1570D}.

A second BNS merger, GW190425, has been discovered in O3a data~\cite{2020ApJ...892L...3A}. This source was more distant and less well localized than GW170817, making the search for an EM counterpart more challenging, and none has been reported.
GW190425 had a total mass $M=3.4^{+0.3}_{-0.1}$~\msun~\cite{2020ApJ...892L...3A}, heavier than other known BNS systems such as binary pulsars, whose total mass has a maximum of $M\sim 2.88$~\msun~\cite{Lazarus2016}. The high total mass of GW190425 is not easy to explain with the formation channels described above.
Formation in an isolated binary does not explain how the system survived the supernovae that generated the neutron stars, because these explosions are more disruptive for more massive progenitor stars.
The alternative, dynamical formation in dense stellar environments seems unlikely: while black holes sink toward the center of the cluster, neutron stars are expected to remain in the less-densely populated outskirts, where they are less likely to meet a companion~\cite{2020ApJ...888L..10Y}. Numerical simulations suggest a merger rate for BNS in dynamical encounters much lower than that inferred from the GW detections\cite{2020ApJ...888L..10Y}.
A possible way to form GW190425 is a secondary object supernova when the binary was already in a tight orbit, too strongly bound to be ripped apart by the explosion. In this scenario, the resulting BNS would merge in only  a few million years. The fast timescale to merger, together with the large orbital acceleration of the neutron stars, could also explain why these systems have not been previously observed by radio surveys~\cite{2020ApJ...892L...3A,Romero-Shaw2020}. Another possibility is that GW190425 was a neutron star-black hole merger~\cite{Kyutoku2020,Foley2020}. 
The interpretation of this event remains a matter of debate.

\section*{Cosmology and  tests of general relativity}

The multi-messenger observations of GW170817 have been used to measure the present value of the Hubble parameter (the Hubble constant)~\cite{Schutz1986,Holz2005}, which parametrizes the local expansion rate of the Universe. Measuring the Hubble constant requires the knowledge of both the redshift  (hence recession speed) and luminosity distance of astrophysical objects. While EM data provides the redshift, it does not usually carry information about the distance. Prior to the detection of GW170817, two main methods have been used to circumvent this obstacle: building a ``cosmic distance ladder'' and using standard candles (objects of known intrinsic brightness); or inferring the expansion rate from the power spectrum of the cosmic microwave background generated in the early Universe.
Those two methods currently yield measurements of the Hubble constant which disagree with each other by more than $5\sigma$~\cite{2019NatAs...3..891V}. Various explanations have been proposed, from systematic errors to new physics~\cite{2019NatAs...3..891V}, while improved measurements have not resolved the discrepancy~\cite{2020MNRAS.tmp.1661W}.

Unlike EM waves, GWs encode information about the luminosity distance of their astrophysical sources~\cite{Schutz1986}, so could resolve the disagreement. For BNS sources for which an EM counterpart is detected, a distance can be derived from the gravitational waves and a redshift from electromagnetic observations of  the host galaxy. For GW170817, this approach provided a measurement of the Hubble constant $H_0=70.0^{+12.0}_{-8.0} \;\rm{km\; s}^{-1}\rm{Mpc}^{-1}$~\cite{2017Natur.551...85A}. While this is still too uncertain to resolve the existing tension, it is limited by the number of detected sources. Predicted detection rates suggest the precision might reach the 1\% level after another five years of GW observationse~\cite{Chen2018}. 

Combining the detections of GW170817 and its associated GRB has provided a constraint on the speed of GWs. Using the time-delay between the arrival of the GRB and the GWs at Earth, it has been inferred that the relative difference between the speed of GWs and the speed of light is in the range $[-3\times 10^{-15},+7\times 10^{-16}]$~\cite{2017ApJ...848L..13A}. This demonstrates that GWs travel at the speed of light, as predicted by GR.
More generally, GWs can be used to test GR, because they probe regions where space-time curvature and gravitational potential are both large~\cite{Will2014}. The phasing evolution of compact binaries can be used to test any non-GR effects that might affect the orbital evolution, e.g. the emission of dipolar radiation, or the propagation of the GWs, which could be caused by the existence of a massive graviton. A measurement of the GWs emitted as a newly formed black hole ``rings'' down to equilibrium, Fig.~\ref{Fig.Combo}B, directly probes dynamics of event horizons.

Numerous tests of GR have been performed using GW data, without finding violations of the theory~\cite{2019PhRvD.100j4036A,O3aTGR}. Some tests search for specific violations, while other unmodeled tests only check whether the detected signals are consistent with GR; neither type show disagreement with the theory.
They have led to an upper limit on the mass of any graviton, $m_g \leq 1.76\times 10^{-23}~\rm{eV}/\rm{c}^2$, which is nearly two times tighter than that obtained with solar system experiments~\cite{O3aTGR}. The heaviest black holes observed during O3a ring down as expected for Kerr black holes in GR~\cite{O3aTGR}. No sign of non-tensorial polarizations, or black hole ``mimickers'' -- exotic compact objects such as boson stars~\cite{2012LRR....15....6L} -- have been found~\cite{O3aTGR}.
GR can explain all the features observed in the GW data.

\section*{The next five years}

LIGO and Virgo are currently being upgraded, and are expected to start their fourth observing run in late 2021 or early 2022~\cite{Aasi:2013wya}. These improved observatories are predicted to detect more than 100 BBHs and many BNSs every year~\cite{Aasi:2013wya}. The global network of gravitational-wave detectors is also due to grow with the addition of the KAGRA detector in Japan~\cite{Akutsu2019} which is also expected to join the fourth observing run, and a third LIGO detector in India~\cite{Iyer2011} which will follow a few years later. Studies are underway to deliver further upgrades after that. The proposed ``A+'' LIGO configuration would provide another factor of {2} in sensitivity~\cite{Aplus}.

More sensitive instruments won't just detect more mergers of binary neutron stars and binary black holes, they will find them at larger distances from Earth, and with higher signal-to-noise ratios.
The improved precision will improve studies of the birth and deaths of compact objects, tests of general relativity, constraints on the behavior of dense nuclear matter, and measurements of the expansion rate of the Universe.
The addition of more detectors leads to improved sky localization, improving the chances of detecting EM counterparts to GW sources and, enabling more systematic studies of the EM emission at all wavelengths, including the energetics and structure of kilonovae.

GW190521 has shown that black holes can exist in the supernovae pair-instability mass gap: the next few years should provide measurements of the mass distribution of black holes in that region, and determine their origin. On the other edge of the mass spectrum, the abundance of massive binary neutron stars such as GW190425 may be determined, and how they form.

The probability of detecting new types of sources will also increase. The most likely is the merger of a neutron star and a black hole, unless GW190814 was such a system. 
Rarer events such as nearby core collapse supernovae are likely to be detectable only if they happen in the Milky Way, but could be observable with gravitational waves, photons and neutrinos. A predicted stochastic background of GW signals produced by thousands of weak unresolved binary mergers is likely to be detected in the near future, complementing the analysis of resolved sources. Continuous near-monochromatic GW emission from Milky Way pulsars might also become observable in the next five years, depending on the ellipticity of neutron stars.

While laser interferometers will continue to explore the explosions and mergers of stellar-mass objects, experiments based on pulsar timing arrays (PTAs), might detect gravitational waves from the supermassive black hole binaries (SMBHs) at the cores of galaxy mergers, providing unique insights on the processes of galaxy formation and evolution.
PTAs use networks of galactic millisecond pulsars as ultra precise clocks. By monitoring the arrival times of the radiopulses, the presence of gravitational waves with frequencies in the nano-Hertz region -- that is, with wavelengths of the order of 10 parsecs --can potentially be detected. Both individual SMBH mergers and the stochastic background generated by unresolved SMBH binaries may be detected. 
Existing PTA results have revealed an excess of power which might be the first hint of the stochastic background GW signal~\cite{Nano12Stoch}.

\section*{Acknowledgments}
I thank the LIGO-Virgo-KAGRA Collaboration, and in particular C. Berry, S. Biscoveanu, M. Fishbach, P. Fritschel, G. Mansell, L. McCuller, R. Weiss and A. Zimmerman for useful comments and discussions during the  preparation of this manuscript. I am also grateful to E. Kara, D. Kaiser and P. Shanahan for comments on a mature version of the manuscript. This is LIGO Document P2000387.
\textbf{Funding} 
I acknowledge support of the National Science Foundation and the LIGO Laboratory. LIGO was constructed by the California Institute of Technology and Massachusetts Institute of Technology with funding from the National Science Foundation and operates under cooperative agreement PHY-0757058. 
\textbf{Competing interests} S.V. is a member of the LIGO Laboratory and the LIGO Scientific Collaboration.
\textbf{Data and materials availability} There are no new data in this review.

\bibliographystyle{Science}
\bibliography{arxivUpdated}

\begin{thebibliography}{100}

\bibitem{Einstein1915}
A.~Einstein, {\it Sitzungsber. K. Preuss. Akad. Wiss. 1, 844\/}  (1915).

\bibitem{Misner2018}
C.~W. Misner, K.~S. Thorne, J.~A. Wheeler, {\it Gravitation\/}, {Princeton
  University Press}, ed. (2018).

\bibitem{Einstein1918}
A.~Einstein, {\it Sitzungsber. K. Preuss. Akad. Wiss. 1, 154\/}  (1916).

\bibitem{Kennefick}
D.~Kennefick, {\it Traveling at the speed of thought: Einstein and the quest
  for Gravitational waves\/}, {Princeton University Press}, ed. (2007).

\bibitem{Weber1969}
J.~Weber, {\it Phys. Rev. Lett.\/} {\bf 22}, 1320 (1969).

\bibitem{1982ApJ...253..908T}
J.~H. {Taylor}, J.~M. {Weisberg}, {\it \apj\/} {\bf 253}, 908 (1982).

\bibitem{Dietrich:2015iva}
T.~Dietrich, S.~Bernuzzi, M.~Ujevic, B.~Br\"ugmann, {\it Phys. Rev. D\/} {\bf
  91}, 124041 (2015).

\bibitem{Weiss1972}
R.~Weiss, {\it Quarterly Report of the Research Laboratory for Electronics, MIT
  Report No. 105\/}  (1972).

\bibitem{2015CQGra..32g4001L}
J.~{Aasi}, {\it et~al.\/}, {\it Classical and Quantum Gravity\/} {\bf 32},
  074001 (2015).

\bibitem{Acernese2015}
F.~Acernese, {\it et~al.\/}, {\it Classical and Quantum Gravity\/} {\bf 32},
  024001 (2014).

\bibitem{Abbott:2016blz}
B.~Abbott, {\it et~al.\/}, {\it \prl\/} {\bf 116}, 061102 (2016).

\bibitem{Aasi:2013wya}
B.~P. {Abbott}, {\it et~al.\/}, {\it Living Reviews in Relativity\/} {\bf 21},
  3 (2018).

\bibitem{2016ApJ...818L..22A}
B.~P. {Abbott}, {\it et~al.\/}, {\it \apjl\/} {\bf 818}, L22 (2016).

\bibitem{TheLIGOScientific:2016pea}
B.~Abbott, {\it et~al.\/}, {\it Phys. Rev. X\/} {\bf 6}, 041015 (2016).

\bibitem{Tse2019}
M.~Tse, {\it et~al.\/}, {\it Phys. Rev. Lett.\/} {\bf 123}, 231107 (2019).

\bibitem{Acernese:2019sbr}
F.~Acernese, {\it et~al.\/}, {\it \prl\/} {\bf 123}, 231108 (2019).

\bibitem{O3aCat}
R.~{Abbott}, {\it et~al.\/}, {\it arXiv:2010.14527\/}  (2020).

\bibitem{Venumadhav2020}
T.~Venumadhav, B.~Zackay, J.~Roulet, L.~Dai, M.~Zaldarriaga, {\it Phys. Rev.
  D\/} {\bf 101}, 083030 (2020).

\bibitem{Zackay2019}
B.~Zackay, T.~Venumadhav, L.~Dai, J.~Roulet, M.~Zaldarriaga, {\it Phys. Rev.
  D\/} {\bf 100}, 023007 (2019).

\bibitem{Nitz2019}
A.~H. Nitz, {\it et~al.\/}, {\it \apj\/} {\bf 872}, 195 (2019).

\bibitem{Nitz2020}
A.~H. Nitz, {\it et~al.\/}, {\it \apj\/} {\bf 891}, 123 (2020).

\bibitem{Reitze}
{LIGO-Virgo Collaboration}, {\it {https://dcc.ligo.org/LIGO-G1901322/public}\/}
   (2020).

\bibitem{1976ApJ...207..574S}
L.~L. {Smarr}, R.~{Blandford}, {\it \apj\/} {\bf 207}, 574 (1976).

\bibitem{Kalogera2007}
V.~Kalogera, K.~Belczynski, C.~Kim, R.~W. O'Shaughnessy, B.~Willems, {\it Phys.
  Rep.\/} {\bf 442}, 75 (2007).

\bibitem{Phinney1991}
E.~S. {Phinney}, S.~{Sigurdsson}, {\it \nat\/} {\bf 349}, 220 (1991).

\bibitem{Benacquista2013}
M.~J. {Benacquista}, J.~M.~B. {Downing}, {\it Living Reviews in Relativity\/}
  {\bf 16}, 4 (2013).

\bibitem{Kroupa2002}
P.~{Kroupa}, {\it Science\/} {\bf 295}, 82 (2002).

\bibitem{Woosley:2016hmi}
S.~E. {Woosley}, {\it \apj\/} {\bf 836}, 244 (2017).

\bibitem{1998ApJ...499..367B}
C.~D. {Bailyn}, R.~K. {Jain}, P.~{Coppi}, J.~A. {Orosz}, {\it \apj\/} {\bf
  499}, 367 (1998).

\bibitem{2011ApJ...741..103F}
W.~M. {Farr}, {\it et~al.\/}, {\it \apj\/} {\bf 741}, 103 (2011).

\bibitem{2010ApJ...725.1918O}
F.~{{\"O}zel}, D.~{Psaltis}, R.~{Narayan}, J.~E. {McClintock}, {\it \apj\/}
  {\bf 725}, 1918 (2010).

\bibitem{LVC02Pop}
B.~P. {Abbott}, {\it et~al.\/}, {\it \apjl\/} {\bf 882}, L24 (2019).

\bibitem{O3aPop}
R.~{Abbott}, {\it et~al.\/}, {\it arXiv:2010.14533\/}  (2020).

\bibitem{May21Det}
R.~{Abbott}, {\it et~al.\/}, {\it \prl\/} {\bf 125}, 101102 (2020).

\bibitem{2020ARAA..58..257G}
J.~E. {Greene}, J.~{Strader}, L.~C. {Ho}, {\it \araa\/} {\bf 58}, 257 (2020).

\bibitem{May21Astro}
R.~{Abbott}, {\it et~al.\/}, {\it \apjl\/} {\bf 900}, L13 (2020).

\bibitem{LVCGW190814}
R.~{Abbott}, {\it et~al.\/}, {\it \apjl\/} {\bf 896}, L44 (2020).

\bibitem{Vitale2014}
S.~{Vitale}, R.~{Lynch}, J.~{Veitch}, V.~{Raymond}, R.~{Sturani}, {\it \prl\/}
  {\bf 112}, 251101 (2014).

\bibitem{Purrer2016}
M.~{P{\"u}rrer}, M.~{Hannam}, F.~{Ohme}, {\it \prd\/} {\bf 93}, 084042 (2016).

\bibitem{2008PhRvD..78d4021R}
{\'E}.~{Racine}, {\it \prd\/} {\bf 78}, 044021 (2008).

\bibitem{Note}
However a highly spinning BBH has been reported~\cite{Zackay2019}, though the
  significance~\cite{2020MNRAS.tmp.2408A} and
  parameters~\cite{2020PhRvD.102j3024H} of the source are still debated.

\bibitem{Fuller2019}
J.~{Fuller}, L.~{Ma}, {\it \apjl\/} {\bf 881}, L1 (2019).

\bibitem{Rodriguez:2016avt}
C.~L. {Rodriguez}, C.-J. {Haster}, S.~{Chatterjee}, V.~{Kalogera}, F.~A.
  {Rasio}, {\it \apjl\/} {\bf 824}, L8 (2016).

\bibitem{Apostolatos1994}
T.~A. {Apostolatos}, C.~{Cutler}, G.~J. {Sussman}, K.~S. {Thorne}, {\it \prd\/}
  {\bf 49}, 6274 (1994).

\bibitem{2020arXiv200408342T}
R.~{Abbott}, {\it et~al.\/}, {\it \prd\/} {\bf 102}, 043015 (2020).

\bibitem{2019MNRAS.484.4216R}
J.~{Roulet}, M.~{Zaldarriaga}, {\it \mnras\/} {\bf 484}, 4216 (2019).

\bibitem{Rodriguez2020}
C.~L. {Rodriguez}, {\it et~al.\/}, {\it \apjl\/} {\bf 896}, L10 (2020).

\bibitem{Mandel2020}
I.~{Mandel}, T.~{Fragos}, {\it \apjl\/} {\bf 895}, L28 (2020).

\bibitem{2020arXiv200504243G}
D.~{Gerosa}, S.~{Vitale}, E.~{Berti}, {\it \prl\/} {\bf 125}, 101103 (2020).

\bibitem{2020ApJ...900..177K}
C.~{Kimball}, {\it et~al.\/}, {\it \apj\/} {\bf 900}, 177 (2020).

\bibitem{1984SvAL...10..177B}
S.~I. {Blinnikov}, I.~D. {Novikov}, T.~V. {Perevodchikova}, A.~G. {Polnarev},
  {\it Soviet Astronomy Letters\/} {\bf 10}, 177 (1984).

\bibitem{Berger2014}
E.~{Berger}, {\it \araa\/} {\bf 52}, 43 (2014).

\bibitem{Metzger2017}
B.~D. {Metzger}, {\it Living Reviews in Relativity\/} {\bf 20}, 3 (2017).

\bibitem{Abbott2017GW170817}
B.~P. Abbott, {\it et~al.\/}, {\it Phys. Rev. Lett.\/} {\bf 119}, 161101
  (2017).

\bibitem{Abbott:2018wiz}
B.~P. {Abbott}, {\it et~al.\/}, {\it \prx\/} {\bf 9}, 011001 (2019).

\bibitem{2017ApJ...848L..13A}
B.~P. {Abbott}, {\it et~al.\/}, {\it \apjl\/} {\bf 848}, L13 (2017).

\bibitem{MMA}
B.~P. Abbott, {\it et~al.\/}, {\it \apjl\/} {\bf 848}, L12 (2017).

\bibitem{Coulter2017}
D.~A. {Coulter}, {\it et~al.\/}, {\it Science\/} {\bf 358}, 1556 (2017).

\bibitem{2018ApJ...856L..18M}
R.~{Margutti}, {\it et~al.\/}, {\it \apjl\/} {\bf 856}, L18 (2018).

\bibitem{2017ApJ...848L..20M}
R.~{Margutti}, {\it et~al.\/}, {\it \apjl\/} {\bf 848}, L20 (2017).

\bibitem{2017ApJ...848L..17C}
P.~S. {Cowperthwaite}, {\it et~al.\/}, {\it \apjl\/} {\bf 848}, L17 (2017).

\bibitem{2017Sci...358.1565E}
P.~A. {Evans}, {\it et~al.\/}, {\it Science\/} {\bf 358}, 1565 (2017).

\bibitem{2017ApJ...848L..16S}
M.~{Soares-Santos}, {\it et~al.\/}, {\it \apjl\/} {\bf 848}, L16 (2017).

\bibitem{Lehner:2001wq}
L.~Lehner, {\it Class. Quant. Grav.\/} {\bf 18}, R25 (2001).

\bibitem{2008PhRvD..77b1502F}
{\'E}.~{\'E}. {Flanagan}, T.~{Hinderer}, {\it \prd\/} {\bf 77}, 021502 (2008).

\bibitem{2015PhRvD..91f4001T}
K.~{Takami}, L.~{Rezzolla}, L.~{Baiotti}, {\it \prd\/} {\bf 91}, 064001 (2015).

\bibitem{2019ApJ...880L..15M}
B.~{Margalit}, B.~D. {Metzger}, {\it \apjl\/} {\bf 880}, L15 (2019).

\bibitem{1998PhRvC..58.1804A}
A.~{Akmal}, V.~R. {Pandharipande}, D.~G. {Ravenhall}, {\it \prc\/} {\bf 58},
  1804 (1998).

\bibitem{2012SPIE.8443E..13G}
K.~C. {Gendreau}, Z.~{Arzoumanian}, T.~{Okajima}, {\it Space Telescopes and
  Instrumentation 2012: Ultraviolet to Gamma Ray\/} (2012), vol. 8443 of {\it
  Society of Photo-Optical Instrumentation Engineers (SPIE) Conference
  Series\/}, p. 844313.

\bibitem{Abbott2018EOS}
B.~P. {Abbott}, {\it et~al.\/}, {\it \prl\/} {\bf 121}, 161101 (2018).

\bibitem{2016ARA&A..54..401O}
F.~{{\"O}zel}, P.~{Freire}, {\it \araa\/} {\bf 54}, 401 (2016).

\bibitem{Yagi2017}
K.~{Yagi}, N.~{Yunes}, {\it \physrep\/} {\bf 681}, 1 (2017).

\bibitem{2018ApJ...852L..29R}
D.~{Radice}, A.~{Perego}, F.~{Zappa}, S.~{Bernuzzi}, {\it \apjl\/} {\bf 852},
  L29 (2018).

\bibitem{Raaijmakers:2019dks}
G.~{Raaijmakers}, {\it et~al.\/}, {\it \apjl\/} {\bf 893}, L21 (2020).

\bibitem{2019ApJ...887L..24M}
M.~C. {Miller}, {\it et~al.\/}, {\it \apjl\/} {\bf 887}, L24 (2019).

\bibitem{2020NatAs...4...72C}
H.~T. {Cromartie}, {\it et~al.\/}, {\it Nature Astronomy\/} {\bf 4}, 72 (2020).

\bibitem{Kasliwal2017}
M.~M. {Kasliwal}, {\it et~al.\/}, {\it Science\/} {\bf 358}, 1559 (2017).

\bibitem{2017Sci...358.1570D}
M.~R. {Drout}, {\it et~al.\/}, {\it Science\/} {\bf 358}, 1570 (2017).

\bibitem{2020ApJ...892L...3A}
B.~P. {Abbott}, {\it et~al.\/}, {\it \apjl\/} {\bf 892}, L3 (2020).

\bibitem{Lazarus2016}
P.~{Lazarus}, {\it et~al.\/}, {\it \apj\/} {\bf 831}, 150 (2016).

\bibitem{2020ApJ...888L..10Y}
C.~S. {Ye}, {\it et~al.\/}, {\it \apjl\/} {\bf 888}, L10 (2020).

\bibitem{Romero-Shaw2020}
I.~M. {Romero-Shaw}, N.~{Farrow}, S.~{Stevenson}, E.~{Thrane}, X.-J. {Zhu},
  {\it \mnras\/} {\bf 496}, L64 (2020).

\bibitem{Kyutoku2020}
K.~{Kyutoku}, {\it et~al.\/}, {\it \apjl\/} {\bf 890}, L4 (2020).

\bibitem{Foley2020}
R.~J. {Foley}, {\it et~al.\/}, {\it \mnras\/} {\bf 494}, 190 (2020).

\bibitem{Schutz1986}
B.~F. {Schutz}, {\it \nat\/} {\bf 323}, 310 (1986).

\bibitem{Holz2005}
D.~E. {Holz}, S.~A. {Hughes}, {\it \apj\/} {\bf 629}, 15 (2005).

\bibitem{2019NatAs...3..891V}
L.~{Verde}, T.~{Treu}, A.~G. {Riess}, {\it Nature Astronomy\/} {\bf 3}, 891
  (2019).

\bibitem{2020MNRAS.tmp.1661W}
K.~C. {Wong}, {\it et~al.\/}, {\it \mnras\/} {\bf 498}, 1420 (2019).

\bibitem{2017Natur.551...85A}
B.~P. {Abbott}, {\it et~al.\/}, {\it \nat\/} {\bf 551}, 85 (2017).

\bibitem{Chen2018}
H.-Y. {Chen}, M.~{Fishbach}, D.~E. {Holz}, {\it \nat\/} {\bf 562}, 545 (2018).

\bibitem{Will2014}
C.~M. {Will}, {\it Living Reviews in Relativity\/} {\bf 17}, 4 (2014).

\bibitem{2019PhRvD.100j4036A}
B.~P. {Abbott}, {\it et~al.\/}, {\it \prd\/} {\bf 100}, 104036 (2019).

\bibitem{O3aTGR}
R.~{Abbott}, {\it et~al.\/}, {\it arXiv:2010.14529\/}  (2020).

\bibitem{2012LRR....15....6L}
S.~L. {Liebling}, C.~{Palenzuela}, {\it Living Reviews in Relativity\/} {\bf
  15}, 6 (2012).

\bibitem{Akutsu2019}
T.~{Akutsu}, {\it et~al.\/}, {\it Nature Astronomy\/} {\bf 3}, 35 (2019).

\bibitem{Iyer2011}
B.~Iyer, {\it et~al.\/}, {\it https://dcc.ligo.org/LIGO-M1100296/public\/}
  (2011).

\bibitem{Aplus}
L.~Barsotti, L.~McCuller, M.~Evans, P.~Fritschel, {\it
  https://dcc.ligo.org/LIGO-T1800042/public\/}  (2018).

\bibitem{Nano12Stoch}
Z.~{Arzoumanian}, {\it et~al.\/}, {\it \apjl\/} {\bf 905}, L34 (2020).

\bibitem{2020MNRAS.tmp.2408A}
G.~Ashton, E.~Thrane, {\it \mnras\/} {\bf 498}, 1905 (2020).

\bibitem{2020PhRvD.102j3024H}
Y.~{Huang}, {\it et~al.\/}, {\it \prd\/} {\bf 102}, 103024 (2020).

\end{thebibliography}

\end{document}